\documentclass[aps,prb,twocolumn,superscriptaddress]{revtex4}

\usepackage{graphicx}
\usepackage{textcomp}
\usepackage{amsmath}
\usepackage{bm}
\usepackage[small]{subfigure}
\graphicspath{{Figures/}}

\begin{document}


\title{Coherence of nitrogen-vacancy electronic spin ensembles in diamond}



\author{P. L. Stanwix}
\affiliation{Harvard-Smithsonian Center for Astrophysics, Cambridge, Massachusetts 02138, USA}
\affiliation{School of Physics, University of Western Australia, Crawley, Western Australia 6009, Australia}
\author{L. M. Pham}
\affiliation{School of Engineering and Applied Sciences, Harvard University, Cambridge, Massachusetts 02138, USA}
\author{J. R. Maze}
\affiliation{Physics Department, Harvard University, Cambridge, Massachusetts 02138, USA}
\affiliation{Facultad de F\'{i}sica, Pontificia Universidad Cat\'{o}lica de Chile, Casilla 306 Santiago, Chile}
\author{D. Le Sage}
\affiliation{Harvard-Smithsonian Center for Astrophysics, Cambridge, Massachusetts 02138, USA}
\author{T. K. Yeung}
\affiliation{School of Engineering and Applied Sciences, Harvard University, Cambridge, Massachusetts 02138, USA}
\author{\\P. Cappellaro}
\affiliation{\mbox{Department of Nuclear Science and Engineering Massachusetts Institute of Technology, Cambridge, Massachusetts 02139, USA}}
\author{P. R. Hemmer}
\affiliation{\mbox{Department of Electrical and Computer Engineering, Texas A\&M University, College Station, Texas 77843, USA}}
\author{A. Yacoby}
\affiliation{Physics Department, Harvard University, Cambridge, Massachusetts 02138, USA}
\author{M. D. Lukin}
\affiliation{Physics Department, Harvard University, Cambridge, Massachusetts 02138, USA}
\author{R. L. Walsworth}
\email[Electronic address: ]{rwalsworth@cfa.harvard.edu}
\affiliation{Harvard-Smithsonian Center for Astrophysics, Cambridge, Massachusetts 02138, USA}
\affiliation{Physics Department, Harvard University, Cambridge, Massachusetts 02138, USA}


\date{\today}

\begin{abstract}
We present an experimental and theoretical study of electronic spin decoherence in ensembles of nitrogen-vacancy (NV) color centers in bulk high-purity diamond at room temperature. Under appropriate conditions, we find ensemble NV spin coherence times ($T_2$) comparable to that of single NVs, with $T_2 > 600 \:\mathrm{\mu s}$ for a sample with natural abundance of $^{13}$C and paramagnetic impurity density $\sim10^{15} \:\mathrm{cm}^{-3}$.  We also observe a sharp decrease of the coherence time with misalignment of the static magnetic field relative to the NV electronic spin axis, consistent with theoretical modeling of NV coupling to a $^{13}$C nuclear spin bath.  The long coherence times and increased signal-to-noise provided by room-temperature NV ensembles will aid many applications of NV centers in precision magnetometry and quantum information.
\end{abstract}

\pacs{}

\maketitle


The Nitrogen-Vacancy (NV) color center in diamond is a promising solid state platform for studying the quantum dynamics of spin systems. Experiments demonstrating spin-state control and manipulation using single NV centers and their nuclear environment~\cite{ChildressScience2006, GaebelNatPhys2006, DuttScience2007, NeumannScience2008, JiangScience2009} have inspired recent work towards quantum information applications of NV diamond, as well as applications to precision magnetic field sensing,~\cite{TaylorNatPhys2008, MazeNature2008, BalasuNature2008} including the life sciences.~\cite{DegenAPL2008, HallPRL2009} Common to all proposals is a desire for long NV electronic spin coherence times, ideally at room temperature, and high signal-to-noise of the NV spin-state-dependent optical fluorescence. 

\begin{figure}[h!]
 \subfigure[]{
 \includegraphics[width=0.42\columnwidth]{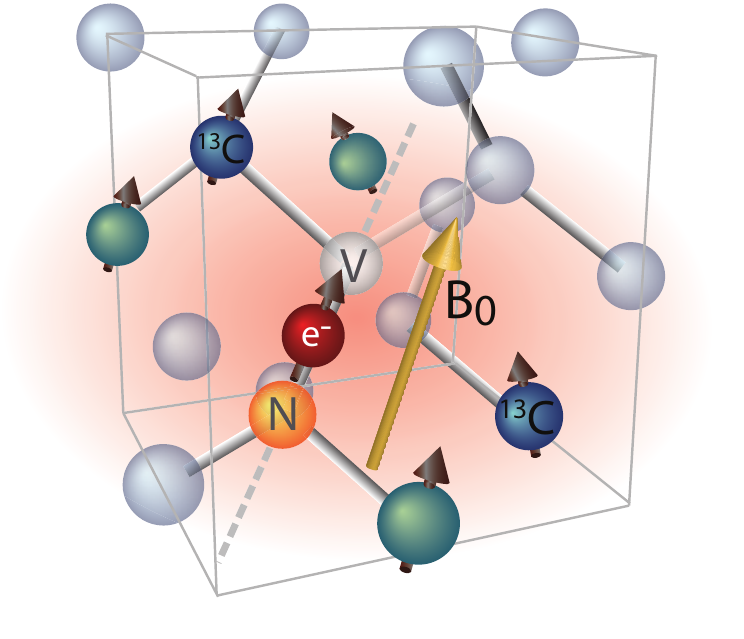}
 \label{fig:NV:crystal}
 }
 \subfigure[]{
 \includegraphics[width=0.45\columnwidth]{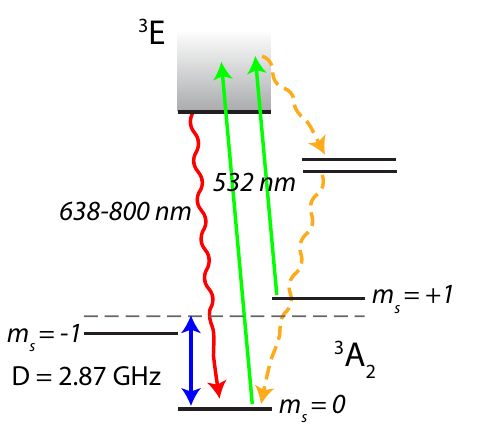}
 \label{fig:NV:levels}
 }
 \subfigure[]{
 \includegraphics[width=0.95\columnwidth]{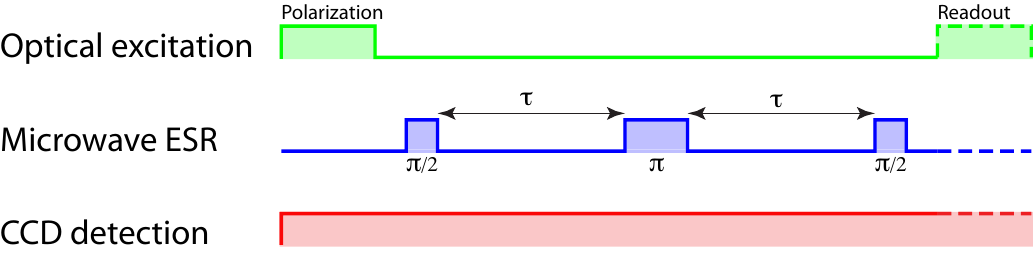}
 \label{fig:NV:seq}
 }
 \label{fig:NV}
 \caption{~\subref{fig:NV:crystal} NV electronic spin axis is defined by nitrogen and vacancy sites, in one of four crystallographic directions. NV orientation subsets in an ensemble can be spectrally selected by applying a static magnetic field, $B_0$. Also shown are $^{13}$C nuclear spins and other paramagnetic impurities such as substitutional nitrogen. ~\subref{fig:NV:levels} NV center electronic energy level structure. Spin polarization and readout is performed by optical excitation and fluorescence detection. Ground state spin manipulation is achieved by resonant microwave excitation. The ground state triplet has a zero magnetic field splitting $D\simeq2.87 \:\mathrm{GHz}$. ~\subref{fig:NV:seq} Hahn-echo experimental sequence used in present measurements.}
 \end{figure}

Electronic spin decoherence of the NV center $\left(S=1 \right)$ is governed by interactions with the surrounding bath of nuclear and paramagnetic spins (Fig.~\ref{fig:NV:crystal}). In high purity diamond, decoherence is dominated by $^{13}$C nuclear spins $\left(I=1/2 \right)$,~\cite{MazePRB2008} which are dispersed through the crystal with a natural abundance of $1.1 \%$. Electronic spin coherence times ($T_2$) longer than $600\: \mathrm{\mu s}$ at room temperature have been observed for individual NV centers in this type of sample.~\cite{MazeNature2008, MizuochiPRB2009} $T_2$ can be increased by isotopically enriching the diamond;~\cite{MizuochiPRB2009} for example, $T_2 > 1.8\:\mathrm{ms}$ at room temperature has been observed in isotopically enriched ultra-pure diamond with $0.3\%$ $^{13}$C.~\cite{BalasuNatMat2009}

In addition to long electronic spin $T_2$, enhancing the signal-to-noise of NV spin-state-dependent fluorescence is of significant benefit to precision magnetometry. To that end, ensembles of NV centers have been proposed,~\cite{TaylorNatPhys2008,AcostaPRB2009} for which signal-to-noise ideally increases as the square root of the number of NVs in the ensemble. The use of large spin ensembles with good coherence times is also important for collective quantum memories.~\cite{ImamoPRL2009,KuboPRL2010,MarcosArXiv2010} Increasing the density of NV centers, however, is accompanied by an increased density of residual nitrogen paramagnetic impurities, which can become the dominant source of NV spin decoherence, reducing $T_2$ below the limit set by $^{13}$C nuclear spins alone.~\cite{VanOortChemPhys1990,VanOortChemPhys1991, HansonScience2008} Furthermore, the large variation of coherence times for individual NV electronic spins, due to the random distribution of $^{13}$C near each NV center, can also affect ensemble coherence. Therefore, a detailed understanding of decoherence mechanisms affecting NV centers -- both single NVs and ensembles -- accompanied by the development of control techniques to mitigate decoherence is of great importance for both precision measurement and quantum information applications.

Only limited studies of electronic spin decoherence in ensembles of NV centers have been previously reported. At room temperature, ensemble $T_2$ times of $58\: \mathrm{\mu s}$ were measured in CVD diamond;~\cite{KennedyAPL2003} whereas at low temperature $\left( < 2 \: \mathrm{K}\right)$ and high magnetic field $\left( > 8 \: \mathrm{T} \right)$, NV ensemble $T_2$ of $250\: \mathrm{\mu s}$ was measured in high-temperature high-pressure type 1b diamond, decreasing to $T_2 < 10 \mathrm{\mu s}$ for temperatures $> 20\:\mathrm{K}$.~\cite{TakahashiPRL2008} 

In the present Rapid communication, we report an experimental and theoretical study of the coherence properties of NV electronic spin ensembles in room temperature diamond samples of different paramagnetic nitrogen (and consequently NV) concentrations. For a lower nitrogen density sample ($\approx 10^{15}\: \mathrm{cm}^{-3}$) with natural abundance of $^{13}$C, we find NV ensemble $T_2$ in excess of $600\: \mathrm{\mu s}$, comparable to the best results for single NV center measurements in natural isotopic abundance diamond and an order of magnitude greater than previous room-temperature ensemble measurements. For a higher nitrogen density sample ($\approx 5\times10^{15}\: \mathrm{cm}^{-3}$) with natural abundance of $^{13}$C, we find an NV ensemble $T_2 \approx 300\: \mathrm{\mu s}$. Furthermore, for both samples we find a sharp decrease in the NV ensemble $T_2$ with misalignment of the static magnetic field relative to the NV electronic spin axis being studied, which is consistent with our theoretical modeling of an ensemble of NV electronic spins interacting with a $^{13}$C spin bath.~\cite{MazePRB2008}

For high-purity diamond, decoherence of single NV electronic spins is dominated by hyperfine interactions with nearby $^{13}$C nuclear spins. The contact term of this interaction decreases exponentially with separation; after a few lattice sites it is not larger than a few MHz.~\cite{GaliPRB2008} Meanwhile, the dipolar part of this interaction decreases as $r^{-3}$ and is responsible for the collapses and revivals observed in Hahn-echo measurements of single NV centers.~\cite{ChildressScience2006} When the externally applied static magnetic field is aligned with the NV axis, the dipolar field contributions of all $^{13}$C nuclei cancel after each $2\pi$ Larmor precession period of the $^{13}$C nuclear spins, as can be measured by the Hahn-echo sequence.~\cite{ChildressScience2006,GaebelNatPhys2006, DuttScience2007} Over longer timescales ($> 600\: \mathrm{\mu s}$ for natural abundance $^{13}$C), weak dipole-dipole interactions between $^{13}$C nuclei and between $^{13}$C and paramagnetic impurities induce fluctuations  (``flip-flops") in the NV--$^{13}$C hyperfine interaction, leading to NV spin decoherence. The predicted form of this decoherence for the Hahn-echo signal of an individual NV is $\exp\left[-\beta \tau^n\right]$, where $n$ is between 3 and 4 and the constant $\beta$ depends on details of the relative location of nearby $^{13}$C nuclei.~\cite{MazePRB2008} However, when the magnetic field makes a small angle $\theta$ with the NV axis, the Larmor precession frequency $\Omega_i$ of individual $^{13}$C nuclei is modified by the hyperfine interaction with the NV center to become dependent on $\theta$ and the relative NV--$^{13}$C position $\bm{r_i}$: $\Omega_i= \Omega_0 + \delta\Omega\left(\theta,\bm{r_i}\right)$. As a consequence, $^{13}$C nuclei do not all precess with the same frequency and thus their modulations of the NV spin coherence do not rephase at the same time in the Hahn-echo sequence, inducing NV spin decoherence of the form $\exp\left[-\alpha\left(\theta\right)b\tau^2\right]$.~\cite{MazePRB2008} Here $\alpha\left(\theta\right)$ describes the misalignment angle dependence of the imperfect Hahn-echo rephasing, and $b$ is proportional to the square of the hyperfine interaction between the NV electronic spin and the nearest $^{13}$C nuclear spin. For small $\theta$, $\alpha\left(\theta\right)\simeq \theta^2$.~\cite{footnote1}

Note that the physics of this angle-dependent NV spin decoherence is fundamentally different from previously observed effects for donor electron spins in semiconductors such as Si:P.~\cite{TyryshkinJPhysCondMatt2006, WitzelPRB2006} For Si:P and similar $S = 1/2$ systems, the electronic spin quantization axis points along the external magnetic field; whereas for NV centers, the anisotropy of the electronic distribution causes spin quantization along the NV axis for small magnetic fields $\left(<500 \:\mathrm{G}\right)$. In Si:P and similar systems, electron spin decoherence is dominated by fluctuations of a dipolar-coupled nuclear spin bath, and is \emph{maximized} when the magnetic field is aligned along the [111] axis, due to the enhanced flip-flop rate of the nuclear spins arising from the angular dependence of their dipolar coupling. As noted above, NV spin decoherence is \emph{minimized} when the magnetic field is aligned with the NV axis.

In diamond samples where the concentration of NV centers is low enough that they do not interact significantly with each other and with nitrogen paramagnetic impurities, the ensemble Hahn-echo signal can be considered as the average of many independent signals from individual NV centers. In this case, the ensemble Hahn-echo signal envelope, $E(\tau)$, can be significantly different from measurements of single NV centers,~\cite{DobrovitskiPRB2008, Abragam1983} because an ensemble contains a broad distribution of spin coherence lifetimes due to variations in the location of $^{13}$C nuclei proximal to individual NVs in the ensemble:

\begin{equation}\label{eq:T2Distr}
E(\tau) = \int db \exp\left(-\alpha\left(\theta\right)b\tau^2\right) f\left(b\right).
\end{equation}

Here, $f\left(b\right)$ is the probability distribution for the magnitude of the NV--proximal $^{13}$C hyperfine-interaction-squared in an ensemble. Depending on details of this distribution, the ensemble Hahn-echo signal envelope can have either a single exponential or Gaussian decay at large times $\tau$.~\cite{DobrovitskiPRB2008, Abragam1983}

 \begin{figure*}[]
 \subfigure[\:Sample A]{
 \includegraphics[width=0.97\columnwidth]{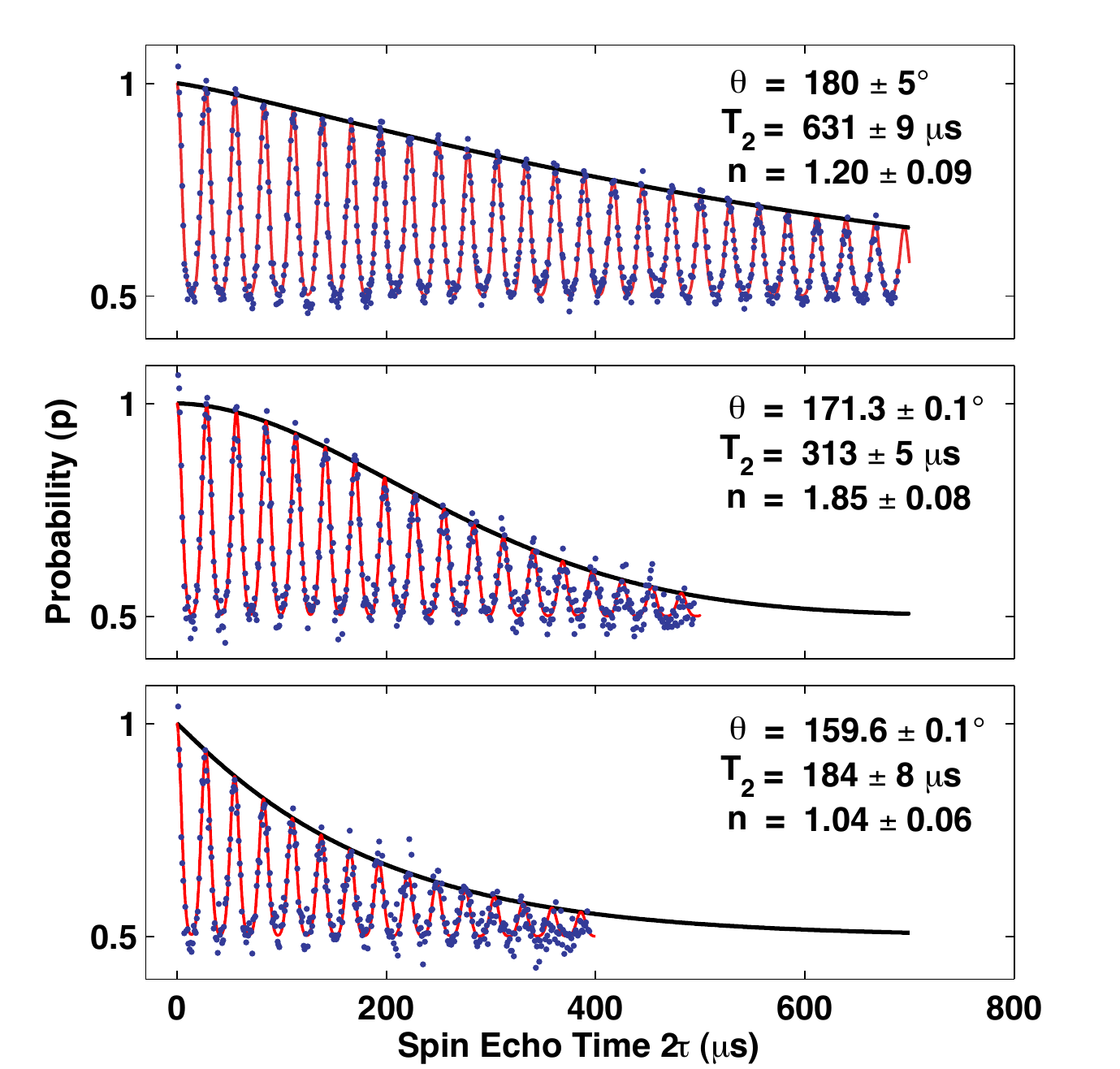}
 \label{fig:T2:bumpy}
 }
 \subfigure[\:Sample B]{
 \includegraphics[width=0.97\columnwidth]{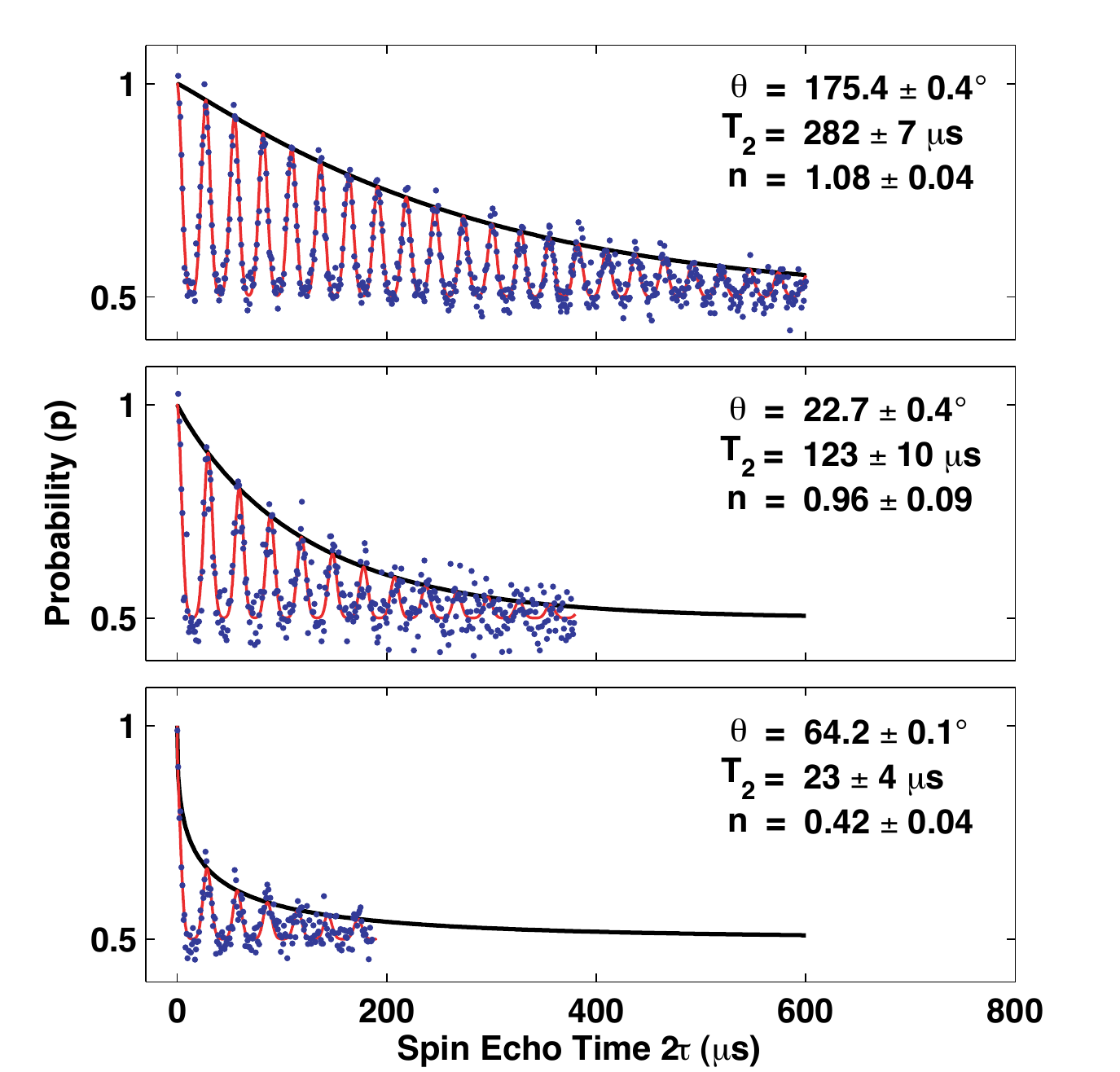}
 \label{fig:T2:wavy}
 }
 \caption{Example Hahn-echo measurements of ensemble NV electronic spin decoherence and dependence on static magnetic field orientation: ~\subref{fig:T2:bumpy} sample A with NV density $\approx3\times10^{13} \: \mathrm{cm}^{-3}$ and N density $\approx 10^{15}\: \mathrm{cm}^{-3}$; and~\subref{fig:T2:wavy} sample B with NV density $\approx1\times10^{14} \: \mathrm{cm}^{-3}$ and N density $\approx 5\times10^{15}\: \mathrm{cm}^{-3}$. Clearly seen in this data is a sharp decrease of the ensemble NV coherence lifetime $T_2$ as the $B_0$ field is misaligned from the NV electronic spin axis, and as the paramagnetic impurity level increases. For each data plot, the left vertical axis indicates the measured probability for NVs in the ensemble to be in the $m_s=0$ state at the end of a Hahn-echo experiment, as a function of the echo time $\tau$; the black line is a fit of $\exp\left[-\left(\tau/_{T_2}\right)^n\right]$ to the overall Hahn-echo signal envelope, where the exponent $n$ is a fit parameter and the long term baseline is taken to be $0.5$ (the mixed spin state); and the errors in $\theta$, $T_2$ and $n$ are given by one-standard-deviation confidence intervals of the fit. The red line is a fit to the observed echo-signal modulations, which are collapses and revivals of NV spin coherence induced by precession of the natural abundance $^{13}$C nuclear spin bath in applied, hyperfine and dipolar magnetic fields.}
 \label{fig:T2}
 \end{figure*}

In our experiments we used a custom-built, wide-field-of-view fluorescence microscope to measure the coherence properties of large ensembles of NV centers. The electronic spin state of negatively charged NV centers can be polarized, manipulated and read-out using optical and microwave excitation (see Figure~\ref{fig:NV:levels}). Spin-state-dependent radiative-relaxation enables NV polarization and readout: NVs initially in $m_s=0$ optically cycle between the ground and an electronically excited ``bright" state, while NVs initially in $|m_s|=1$ have a significant branching to a long lived metastable ``dark" state that non-radiatively relaxes to $m_s=0$. Optical excitation is provided by a switched $532\:\mathrm{nm}$ laser source, focussed onto the sample with a 20x objective. Fluorescence emanating from the sample is collected back through the same objective, then filtered and imaged onto a CCD array. The use of a CCD array allows the sample response to be spatially resolved with resolution $<1 \: \mathrm{\mu m}$ and field-of-view $>100 \: \mathrm{\mu m}$. The ensemble of NVs is coherently manipulated  on a ground-state electronic spin transition with a resonant microwave magnetic field generated from a loop antenna placed close to the diamond sample, which produces a homogenous microwave $B_1$ field over the region of interest. The degeneracy of the $m_s=0$ to $m_s=\pm1$ ground-state spin transitions is lifted by a uniform static $B_0$ magnetic field, applied using a permanent magnet trimmed by a 3-axis set of Helmholtz coils. Precise alignment of $B_0$ allows a subset of NVs in the ensemble, corresponding to one of four crystallographic orientations, to be spectrally distinguished by their spin transitions. The orientation $\theta$ and magnitude $B_0$ of the applied magnetic field relative to the axis of one class of NV centers is determined from the Zeeman shifts of the NV ground-state electronic spin resonances (ESR) together with the measured Hahn-echo revival frequency, $B_0 = \omega_L/\gamma_{_{^{13}C}}$.~\cite{footnote2}

We employed two CVD diamond samples in our measurements of NV ensembles. Each sample consists of a pure diamond substrate ($3\times3\times0.2\:\mathrm{mm}$) with $<1\:\mathrm{ppm}$ of ambient nitrogen impurity and $<10^{12}\: \mathrm{cm}^{-3}$ NV density, which was overgrown with a high NV content layer $\approx2 \:\mathrm{\mu m}$ thick. The density of NV centers in the thin layer region of sample A and B is approximately $3\times10^{13}\:\mathrm{cm}^{-3}$ and $1\times10^{14}\:\mathrm{cm}^{-3}$ respectively, as determined by fluorescence measurements normalized to single NV measurements made on each sample using a confocal microscope optimized for NV measurements, similar to that used in Ref. ~\onlinecite{MazeNature2008}. The density of nitrogen impurities in the samples is approximately 50 times larger, as determined by the CVD growth procedure. A Hahn-echo pulse sequence (illustrated in Fig.~\ref{fig:NV:seq}) was used to measure the electronic spin coherence time of the NV ensemble in each of the two samples. NV centers were first prepared in the $m_s=0$ ground state by applying a $532\:\mathrm{nm}$ laser excitation pulse. The Hahn-echo sequence ($\pi/_2-\tau-\pi-\tau-\pi/_2$) was then repeated for a range of free spin evolution periods $\tau$ (the final $\pi/_2$ pulse in this sequence projects a coherent superposition of NV spin states into a population difference, which is detected via spin-state-dependent fluorescence). The data presented here is the integrated response of regions with area $\approx 30\:\mathrm{\mu m}^2$ and $2\:\mathrm{\mu m}$ depth in each sample, averaged over $\sim10^5$ identical experiments. Thus, NV ensembles consisted of $\sim 10^3$ NV spins of a single crystallographic class in the detection region. Over macroscopic distances ($\sim 1\:\mathrm{mm}$), each diamond sample was found to have uniform spin coherence properties.

 \begin{figure}[h]
 \subfigure[]{
 \includegraphics[width=0.85\columnwidth]{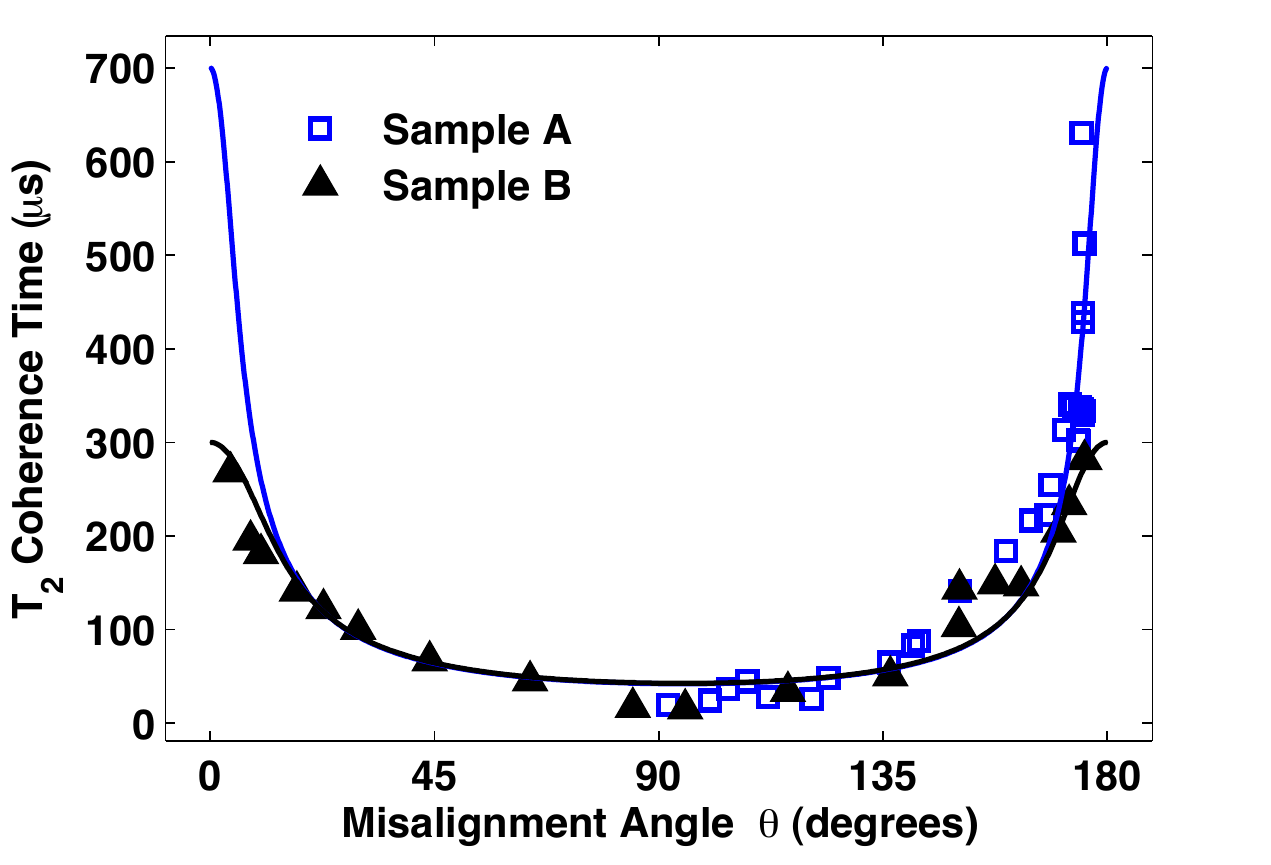}
 \label{fig:orientation:T2}
 }
 \subfigure[]{
 \includegraphics[width=0.85\columnwidth]{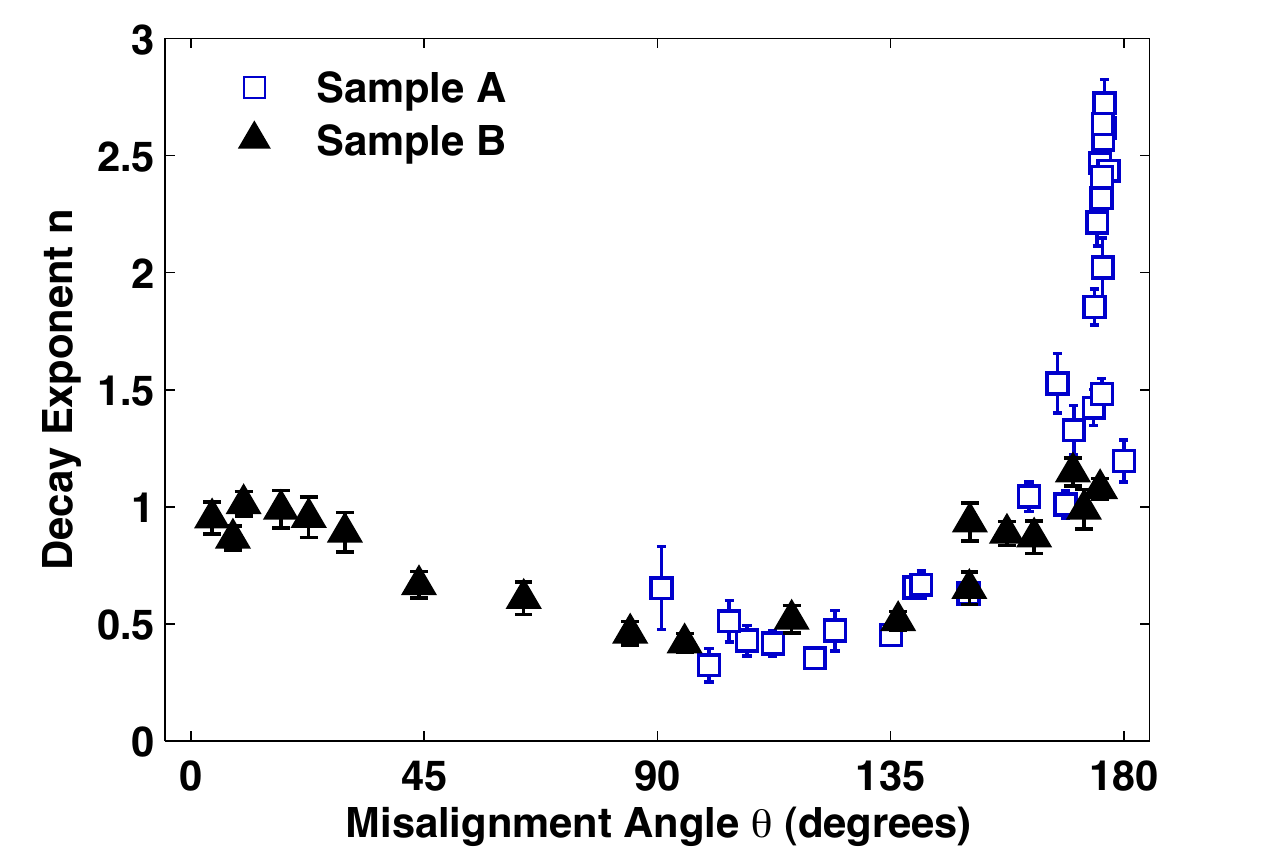}
 \label{fig:orientation:n}
 }
 \caption{Summary of Hahn-echo measurements as a function of misalignment angle $\theta$ between the static magnetic field, $B_0$, and the NV electronic axis:~\subref{fig:orientation:T2} coherence lifetime $T_2$ and~\subref{fig:orientation:n} decay exponent $n$, determined from fits of the signal envelope to $\exp\left[-\left(\tau/_{T_2}\right)^n\right]$.  Sample A (lower NV and nitrogen concentration) was measured for $\theta$ ranging from 90 to 180 degrees; sample B (higher NV and nitrogen concentration) was measured for $\theta$ from 0 to 180 degrees.  Solid curve in~\subref{fig:orientation:T2} is the prediction of the NV ensemble decoherence model, as described in the main text.}
 \label{fig:orientation}
 \end{figure}

As shown in Figs.~\ref{fig:T2} and~\ref{fig:orientation}, we measured ensemble NV coherence times comparable to that for single NVs in each sample: $T_2>600 \:\mathrm{\mu s}$ for sample A and $T_2\approx300 \:\mathrm{\mu s}$ for sample B, with the static magnetic field $B_0$ well aligned with the NV electronic spin axis. Our measurements also confirm the predicted~\cite{MazePRB2008} sharp decrease of NV $T_2$ with small misalignment of $B_0$ from the NV axis. Example Hahn-echo measurements for three representative misalignment angles $\theta$ are shown in Fig.~\ref{fig:T2} for samples A and B; whereas Fig.~\ref{fig:orientation} summarizes the results of a larger set of systematic measurements of ensemble NV $T_2$ as a function of $\theta$. In all these measurements, the Hahn-echo signal envelope was fit to $\exp\left[-\left(\tau/_{T_2}\right)^n\right]$ to determine the coherence lifetime $T_2$ (Fig.~\ref{fig:orientation:T2}) and decay exponent $n$ (Fig.~\ref{fig:orientation:n}). Good agreement is found between our experimentally-determined values for $T_2$ as a function of $\theta$ and theoretical predictions (see solid curve in Fig.~\ref{fig:orientation:T2}), which include the decoherence contributions from proximal $^{13}$C nuclear spins via the form $\exp\left[-\alpha \left(\theta \right) b \tau^2\right]$.~\cite{MazePRB2008,footnote3} Decoherence due to $^{13}$C ``flip-flops" was included phenomenologically by multiplying the simulated NV echo signal by $\exp\left[-\left(\tau/_{T_2\left(\theta=0\right)}\right)^4\right]$. The low values for the decay exponent $n$ in ensemble experiments is explained by the large dispersion of the dominant NV--$^{13}$C dipole-dipole interaction (see Eq.~\ref{eq:T2Distr}). The distribution of this interaction squared, $f\left(b\right)$, was determined by calculating the effect of $^{13}$C nuclei placed randomly in the diamond lattice at natural isotopic abundance. For the dominant NV--$^{13}$C dipole-dipole interaction, we find that this distribution scales as $f\left(b\right)\sim b^{-3/2}$ for $^{13}$C nuclei within about $10\:\mathrm{nm}$ of an NV center.  Thus, $f\left(b\right)$ more closely resembles a Lorentzian than a Gaussian distribution, resulting in Hahn-echo signal envelopes that decay more like a single exponential than a Gaussian at large times (see Fig.~\ref{fig:T2}).  Note that for sample B, the experimentally-determined decay exponent is $n\simeq 1$ even for $\theta \simeq 0$ (Fig.~\ref{fig:orientation:n}), which could result from interactions between NV spins and the higher density of paramagnetic impurities (nitrogen and NVs) in this sample.

In summary, we demonstrated experimentally that large ensembles of NV centers in high-purity diamond with natural abundance of $^{13}$C can have electronic spin coherence lifetimes at room temperature that are comparable to the best measured for single NV centers ($T_2>600 \:\mathrm{\mu s}$).  We also found a sharp decrease in NV $T_2$ as the applied magnetic field is misaligned from the NV axis, consistent with the predictions of our model for an ensemble of NV electronic spins $\left(S=1\right)$ coupled via position-dependent hyperfine interactions to a $^{13}$C nuclear spin bath, leading to imperfect $^{13}$C spin-echo revivals and hence NV decoherence. Our results demonstrate the potential of NV ensembles for applications in precision magnetometry in both the physical and life sciences, combining long electronic spin coherence times at room temperature with the enhanced signal-to-noise ratio provided by many NVs in the detection volume.~\cite{TaylorNatPhys2008} In addition, the demonstrated techniques could be used to increase the coherence time of solid-state quantum memories, such as an NV electronic spin ensemble coupled to a superconducting resonator~\cite{ImamoPRL2009,KuboPRL2010} or flux qubit.~\cite{MarcosArXiv2010}

This work was supported by NIST, NSF and DARPA. We gratefully acknowledge the provision of diamond samples by Apollo Diamond and technical discussions with Patrick Doering and David Glenn.

\bibliography{B0T2e}

\begin{thebibliography}{30}
\bibitem[{foo({\natexlab{b}})}]{footnote2}
\bibinfo{note}{The magnetic field magnitude $B_0$ and angle $\theta$ relative
  to the axis of one class of NVs in the ensemble can be extracted from the ESR
  transition frequencies $\nu_1$ and $\nu_2$ using $B_0=\sqrt{(P-D^2)/3}$ and
  $\cos^2\theta = \frac{Q+9DB^2+2D^3}{27DB^2}$, where $D \simeq
  2.87\:\mathrm{GHz}$ is the zero field splitting, $P
  =\nu_1^2+\nu_2^2-\nu_1\nu_2$ and
  $Q=(\nu_1+\nu_2)(2\nu_1^2+2\nu_2^2-5\nu_1\nu_2)$.}
\bibitem[{foo({\natexlab{c}})}]{footnote3}
\bibinfo{note}{The angle dependence can be modeled by $\alpha = \left(c_1^2 +
  c_2^2\cos2\theta\right)\sin^2\theta$, where $c_1$ and $c_2$ are parameters
  that depend on the hyperfine interaction between an NV center and a proximal
  nuclear spin.}
\end{thebibliography}

\end{document}